\documentclass[conference]{IEEEtran}
\IEEEoverridecommandlockouts
\usepackage{cite}
\usepackage{amsmath,amssymb,amsfonts}
\usepackage{algorithmic}
\usepackage{graphicx}
\usepackage{textcomp}
\usepackage{xcolor}
\usepackage{soul,color}
\def\BibTeX{{\rm B\kern-.05em{\sc i\kern-.025em b}\kern-.08em
    T\kern-.1667em\lower.7ex\hbox{E}\kern-.125emX}}
    
\usepackage[margin=0.75in]{geometry}

\usepackage{algorithm}
\usepackage{algorithmic}

\usepackage{multirow}

\begin{document}

\title{Fed-DDM: A Federated Ledgers based Framework for Hierarchical Decentralized Data Marketplaces}

\author{\IEEEauthorblockN{Ronghua Xu, Yu Chen}
\IEEEauthorblockA{Department of Electrical and Computer Engineering, Binghamton University, Binghamton, NY 13902, USA\\
\{rxu22, ychen\}@binghamton.edu}
}



\maketitle

\begin{abstract}
Data marketplaces (DMs) promote the benefits of the Internet of Things (IoT) in smart cities. To facilitate the easy exchanges of real-time IoT data streams between device owners and third-party applications, it is required to provide scalable, interoperable, and secured services for large numbers of distributed IoT devices operated by different application vendors. Thanks to decentralization, immutability and auditability, Blockchain is promising to enable a tamper-proof and trust-free framework to enhance performance and security issues in centralized DMs. However, directly integrating blockchains into large scale IoT-based DMs still faces many limitations, such as high resource and energy demands, low transactions throughput, poor scalability and challenges in privacy preservation.
This paper introduces a novel Federated Ledgers based Framework for Hierarchical Decentralized Data Marketplaces (Fed-DDM).
In Fed-DDM, participants are divided into multiple permissioned domains given their registrations. Each domain leverages an efficient Byzantine Fault Tolerance (BFT) consensus protocol to commit transactions of domain on a private intra-ledger.
A public inter-ledger network adopts a scalable Proof-of-Work (PoW) consensus protocol to federate multiple private intra-ledger networks.
We design a smart contract enabled inter-ledger protocol to guarantee security of the cross-domain operations on public federated ledger without exposing sensitive privacy information from private ledgers.
A proof-of-concept prototype is implemented, and the experimental results verify the feasibility of the proposed Fed-DDM solution with performance and security guarantees. 

\end{abstract}


\section{Introduction}
\label{sec:intro}
The proliferation of the Internet of Things (IoT) technology allows the concept of Smart Cities to become feasible, and IoT-based applications have significantly improved the lives of their residents by enhancing health, safety, and convenience. However, the first generation of IoT deployment (IoT 1.0) for smart cities has come across several challenges that prevent wider adoption \cite{ramachandran2019towards}. Data marketplaces (DM), including Intelligent IoT Integrator (I3) \cite{krishnamachari2018i3}, Ocean Protocol \cite{oceanpocotol}, and IOTA Data Marketplace \cite{iota}, are being developed to enhance the adoption of IoT in smart communities and smart cities. Such data marketplace initiatives focus on building ``data rivers'' that allow data streams from different entities to be merged, analyzed, processed, and acted upon as needed to support a diverse set of applications \cite{krishnamachari2018i3}. In a commercial aspect, the data marketplace allows a large volume of device owners to sell data streams, while different applications can buy one or more data streams to develop applications for smart communities. 

While the DMs facilitate the easy exchange of real-time IoT data streams between device owners and third-party applications through the marketplace middleware, it also brings new architecture and security concerns. The conventional DMs rely on a centralized platform where all user data flows are merged and then traded among participants. Such a centralized data server can be a performance bottleneck and susceptible to a single point of failure risk. In addition, as a major requirement for DMs, data quality should be ensured in terms of data integrity and auditability. In other words, the data should be consistent, unaltered and auditable through the entire lifetime. Furthermore, DMs also need incentive mechanisms that encourage more buyers and sellers to join marketplaces and it should also provide mechanisms to prevent dishonest behaviors, for example, through policies that punish the malicious actor.

Evolving from the distributed ledger technology (DLT), blockchain has demonstrated a great potential to revolutionize information technologies (IT). Blockchain relies on a decentralized peer-to-peer (P2P) consensus network to ensure inherent security guarantees rather than centralized trusted third-party authorities. Such a decentralization architecture mitigates the bottleneck performance and removes the single point of failure issue, which were inherent in a central hub, like data aggregation in DMs. Also, blockchains leverage consensus protocols and DLTs to support immutability, auditability and traceability for data provenience, and it is essential to ensure the quality of data in use and storage in DMs. Furthermore, incentive mechanism in blockchains motivates more users to actively participate in data trading to gain benefits while discouraging dishonest participants from misbehaving via penalties. 

\subsection{Challenges in Blockchain based DMs}
To enable an efficient, scalable and secure mechanism for IoT-based DMs, shifting from centralized paradigms to blockchain-based methods has been a hot research topic. Previously reported works \cite{oceanpocotol, iota, ramachandran2019trinity, ramachandran2018towards, xu2019blendsm} considered adopting blockchain to support decentralized DMs (DDM) by merely integrating existing cryptocurrency oriented blockchain technologies into IoT-based DMs. However, these approaches rely on mono-chain architectures which also incur tremendous challenges in terms of performance, scalability and security. 

As IoT devices are strictly constrained by computation and storage requirements, such that Proof-of-Work (PoW) \cite{nakamoto2019bitcoin} algorithm and its variants are not affordable. Compared with PoW with high energy consumption and low throughput, Practical Byzantine Fault Tolerance (PBFT) \cite{castro1999practical} demonstrates high throughout, lower latency and limited overhead, but it only allows a very limited network scalability in terms of the number of validators. Therefore, it is very difficult for a monolithic chain based DDM system to handle the blockchain trilemma \cite{zhou2020solutions}, which points out that decentralization, scalability and security cannot perfectly co-exist.

Another challenge in designing an efficient DDM framework lies in heterogeneous IoT networks. Owing to dynamics and non-standard technologies in IoT systems, large-scale practical DMs are likely to use IoT devices from fragmented domains, which are managed by different vendors to improve management complexity. To ensure performance and security, the administrator needs to accurately monitor the domain system's performance and audit the behavior of participants or devices to identify potentially malicious activities \cite{tseng2020blockchain}. However, domain and devices owner could have various privacy preserving requirements, thus, directly recording sensitive information into transactions that are committed on public ledger may leak user's privacy. 


\subsection{Goals and Contributions}

This paper proposes Fed-DDM, a novel federated ledgers based framework for hierarchical decentralized data marketplaces. Unlike existing mono-chain based DDMs solutions, our Fed-DDM adopts federated networking framework \cite{xu2018federated} and hierarchical cross-chain \cite{cosmos} consensus mechanism to achieve high scalability and ensure security and privacy-preservation of inter-domain transactions. 

In Fed-DDM, each domain relies on a private intra-ledger to record transactions among its members. As each domain is a permissioned network, all data on the intra-ledger is only accessed by authorized members within the domain. Given permissioned network of a domain, an efficient Byzantine Fault Tolerance (BFT) \cite{lamport1982byzantine} based consensus protocol can be executed by validators to achieve low latency and high throughput of transactions in the synchronous network environment. Such a private blockchain managed by the local domain aims to support partial decentralization with performance and privacy preserving guarantees.

Meanwhile, multiple private intra-ledger networks are federated through a high-level and public inter-ledger network, which uses a scalable PoW consensus protocol to secure cross-domain operations under an asynchronous network environment. For cross-domain operations, our inter-ledger transactions only include checkpoints information rather than raw transactions on private intra-ledgers. Therefore, a smart contract enabled inter-ledger protocol can guarantee auditability, immutability and provenance of cross-domain operations without exposing sensitive data of private ledgers managed by domains. Such a federated ledger structure ensures scalability and security without sacrificing performance and privacy requirements of individual domains. 

The remainder of this paper is organized as follows: Section \ref{sec:relatedwork} discusses the state-of-the-art research on DDMs briefly, then reviews related work on federated ledgers. Section \ref{sec:architecture} introduces the rationale and architecture of Fed-DDM, and a novel smart contract enabled inter-ledger transaction protocol is explained in Section \ref{sec:crossledger}. Section \ref{sec:experiment} presents prototype implementation with numerical results, and discusses performance improvements and security insurances. Section \ref{sec:conclusion} concludes the paper with the future work. 

\section{State of The Art and Related Work}
\label{sec:relatedwork}
\subsection{Decentralized Data Marketplaces}
Blockchain \cite{nakamoto2019bitcoin} has demonstrated great potential to revolutionize the fundamentals of information technology (IT) due to many attractive properties, such as decentralization, immutability, and transparency. Decentralized Application (DApp), which is built in the form of a smart contract and deployed on a blockchain network, performs pre-defined algorithms and agreements without relying on a third-party intermediary. Blockchain and smart contracts together provide a decentralized solution to address issues in data marketplaces \cite{xu2019blendsm}.

An earlier work has explored how a DDM could be created using blockchain and other distributed ledger technologies \cite{ramachandran2018towards}. The potential benefits of such a decentralized architecture are highlighted, and the preliminary work shows how key elements of such a decentralized marketplace could be implemented using smart contracts. Inspired by blockchain and containerized microservices technology, BlendSM-DDM \cite{xu2019blendsm} is proposed as a decentralized microservice architecture for the DDM system, which is mainly to secure data exchange among different device owners and application developers. The BlendSM-DDM decouples DDM applications and security services into multiple containerized microservices rather than using a monolithic service architecture. It supports loose-coupling, fine-granularity and easy-maintenance for DDM systems.

Other marketplaces include Ocean Protocol \cite{oceanpocotol}, IOTA Data Marketplace \cite{iota} and Trinity \cite{ramachandran2019trinity}. Ocean is a decentralized protocol and network of artificial intelligence (AI) data/services. Ocean utilizes a smart contract called Service Execution Agreement (SEA) to handle each step in the directed acyclic graph (DAG) of computing and storage, and these SEA contracts are deployed on an Ethereum network to achieve a decentralized orchestration. IOTA is a cryptocurrency designed for the IoT industry. IoTA tangle, which is a directed acyclic graph (DAG) for storing transactions, provides a secure data communication protocol and zero fee micro-transaction system for the IoT/M2M. Trinity \cite{ramachandran2019trinity} distributes the pub-sub broker using blockchain technology to prevent byzantine faults and relies on an immutable distributed ledger to ensure data persistence. Trinity is implemented using a Message Queuing Telemetry Transport (MQTT) broker running on a pluggable blockchain platform that supports multiple existing blockchain networks, like Tendermint. Trinity consumes minimal resources on IoT devices and guarantees ordering, fault-tolerance and persistence across trust boundaries.

Above mentioned DDM solutions are merely integrating existing cryptocurrency oriented blockchain technologies into IoT-based data marketplaces scenarios. Unavoidably they face critical challenges. Most of the earlier efforts rely on computation-intensive proof-of-work (PoW) consensus mechanisms to guarantee the network scalability and mitigate Sybil attacks \cite{oceanpocotol, ramachandran2018towards, xu2019blendsm} at the cost of low throughput and high energy consumption. Trinity \cite{ramachandran2019trinity} and IOTA \cite{iota} demonstrate better performance in terms of throughput, latency, and overhead. However, they either rely on identity authentication or allow very limited network scalability. 

Unlike the reported approaches, our Fed-DDM is based on a federated ledger framework: a scalable cross-chain mechanism that provides decentralization, immutability and auditability for hierarchical DDM applications. Compared with Trinity, which is a blockchain-enabled publish-subscribe system, Fed-DDM is aimed to build a scalable, efficient, security and privacy-preserving DDM platform.

\subsection{Related Work}
The Fed-DDM system is built on top of several key ideas from prior work on federated networks and cross-chain protocols, as described below.

In general, the PoW-based Nakamoto blockchain \cite{nakamoto2019bitcoin} provides good scalability and probabilistic finality at the cost of lower throughput and high energy consumption. Compared to PoW style blockchains, BFT based blockchain networks offer excellent performance and a deterministic finality but demonstrate limited scalability. Adopting relay technique to connect different blockchains together, cross-chain approaches \cite{cosmos, xu2020blendsps, tseng2020blockchain, nguyen2021fedchain, kan2018multiple, ahmad2019blocktrail} expect to build a large-scale blockchain network and ensure inter-operability among multiple chains. Therefore, cross-chain mechanism provides a prospective solution to scale out blockchain systems without sacrificing performance and security guarantees.

Cosmos \cite{cosmos} is a novel blockchain network architecture, which uses a Cosmos Hub to connect multiple independent blockchains, called zones. The Cosmos Hub is the the first zone on Cosmos and relies on a multi-asset proof-of-stake (PoS) cryptocurrency as a consensus network. While other zones are powered by a secure PBFT-like consensus protocol, like Tendermint \cite{kwon2014tendermint}, to ensure high-performance, consistency and fork accountability guarantees. The Cosmos Hub and zones leverage an inter-blockchain communication (IBC) protocol to communicate with each other. Therefore, Cosmos Hub acts as a secured relay intermediate to support inter-operability between zones. Unlike Cosmos, our Fed-DDM uses PoW-like Ethereum and smart contract to build a federated ledger, which not only supports cryptocurrency-based payments, but also brings programmability to achieve non-cash transactions and complicated business logic. 

Similar to Cosmos, an interactive multiple blockchain architecture is proposed \cite{kan2018multiple} to support exchanging information across arbitrary blockchain systems. An inter-blockchain connection model is designed for routing management and message transferring, and a cross-chain transaction protocol aims to guarantee atomicity and consistency. A router blockchain acts as an intermediate hub to transfer those transactions among private blockchains. However, privacy preservation is not considered in the cross-chain transactions scenario. In our Fed-DDM, only checkpoints of private ledgers are included in transactions for inter-ledger verifications. Thus, privacy of raw data on private ledger is guaranteed. 

To address issues of single ledger in terms of increasing the storage footprint and bottleneck of parallel processing, BlockTrail \cite{ahmad2019blocktrail} leverages the hierarchical structure to enable a scalable and efficient multilayer blockchain solution for auditing applications. BlockTrail provides a theoretical analysis on performance and security, but cross-chain transactions processing is not discussed. To address challenges in integrating blockchain and heterogeneous IoT systems, a hierarchical blockchain architecture \cite{tseng2020blockchain} is proposed for IoT system. Each IoT subsystem is secured by its private chain, called core IoT, and a public blockchain connects several core IoTs to enable data exchange, coordination, and other important features of the whole system. Similar to the hierarchical architecture, a BlendSPS \cite{xu2020blendsps} solution is proposed to enable decentralized security services for multi-domain applications in smart public safety systems. Unlike above mentioned hierarchical blockchains, our Fed-DDM leverages a federated security mechanism and supports a privacy-preserving inter-ledger transaction protocol. 

\section{Fed-DDM: Rationale and Architecture}
\label{sec:architecture}
\subsection{Prerequisites}
A DDM system could get fragmented or be hard to scale owing to system management complexity. Therefore, Fed-DDM adopts a hierarchical network paradigm consisting of a public federated network and multiple private domain networks. Each domain relies on a permissioned network management, assuming that the domain administrator is a trust oracle to maintain identity profiles for all valid nodes. Such a permissioned network provides basic security primitives, like public key infrastructure (PKI), identity authentication and access control, etc.. Following the idea of delegation, each domain chooses a small subset of the nodes within its network to jointly participate the public federated network for cross-domain data services and coordination.

We assume a synchronous network environment for each domain network such that all messages could be delivered successfully within a fixed upper bounded delay. The intra-ledger of a domain network is based on a BFT consensus protocol to achieve a lower latency and a higher transaction throughput. The public inter-ledger relies on a asynchronous network assumption that the upper bounded delay for delivering messages is not exist. Thus, the high-level inter-ledger network is based on a PoW consensus protocol to ensure global security and guarantee scalability.  

Regarding heterogeneous platforms of IoT systems, keeping a rigid monolithic service-oriented architecture (SOA) could also hurt scalability and flexibility of system. The virtualization technology, like virtual machines (VMs) or containers, is platform-independent and could provide resource abstraction and isolation features \cite{nagothu2018microservice}. Thus, Fed-DDM leverages a lightweight Docker \cite{docker} container technology to implement microservices-based SOA for solving the heterogeneity challenge in IoT-enabled DDMs. 

\subsection{System Overview}
\begin{figure} [t]
\begin{center}
\begin{tabular}{c}
\includegraphics[height=12.5 cm]{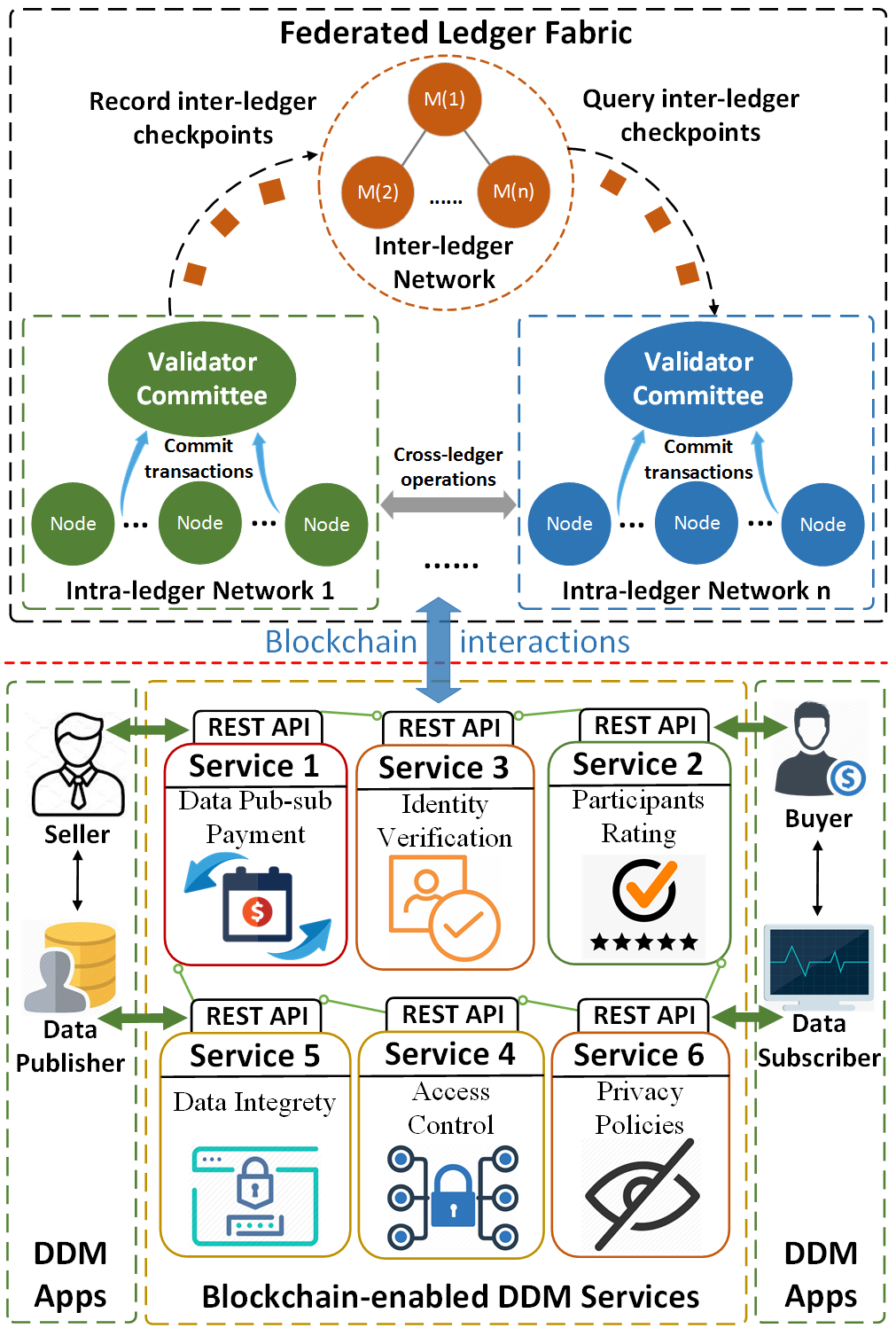}
\end{tabular}
\end{center}
\vspace{-12pt}
\caption[example] { \label{fig:Fed-DDM_arch} Illustration of Fed-DDM system architecture.}
\vspace{-15pt}
\end{figure}

Figure \ref{fig:Fed-DDM_arch} shows the Fed-DDM architecture consisting of (i) a \emph{blockchain-enabled DDM services} framework that leverages microservices to support flexible, efficient and secure DDM business functions, and (ii) a \emph{federated ledger fabric} that provides a fundamental networking and decentralized infrastructure to enhance security and privacy preserving properties for the hierarchical DDM.

\subsubsection{Blockchain-enabled DDM services} as the lower part of Fig. \ref{fig:Fed-DDM_arch} shows, these services are considered as an intermediate services layer to bridge upper-level DDM apps and underlying blockchain network. The key elements and security features are decoupled into multiple microservices and deployed on distributed computing hosts. Following the containerized microservices architecture, each functional unit is computationally affordable on host nodes, especially for those resource-constrained IoT devices. The fine-granularity and loose-coupling features of architecture allow fast development and easy deployment among different application vendors using non-standard development. Each microservice exposes a set of RESTful web-service APIs to accept service requests sent by DDM apps and uses its local ABIs to interact with SCs or access data on blockchains. 

In DDM systems, all users (sellers and buyers) and brokers (data publishers and subscribers) must finish registration for a blockchain network, then use their blockchain account addresses as unique IDs to join the data exchange and payment activity. In general, users are light nodes that only store accounts related information and do not join consensus or mining tasks. While brokers are full nodes that enforce all operations in blockchain, like transactions propagation and blocks verification, etc.. 

As two basic business functions, the data payment service supports token or cryptocurrency based micro-payment protocol, and rating service is mainly to evaluate the quality of the data product by a seller and reliability of payment by a buyer. Identity verification service relies on a virtual trust zone method \cite{xu2019exploration} to enable a decentralized identity authorization. Access control service uses a fine-grained capability-based access control mechanism \cite{xu2018blendcac} to support decentralized access authorization and verification. Through a decentralized data integrity scheme \cite{nikouei2018real}, data integrity aims to secure data sharing among trust-less participants in Fed-DDM. To protect privacy-sensitive data, secure communications using encryption and saving a hash value of sensitive data are feasible solutions. Furthermore, properly access control policy design could effectively prevent against unauthorized access to privacy data in cross-domain scenarios. 

\subsubsection{Federated ledger fabric}
the upper part of Fig. \ref{fig:Fed-DDM_arch} demonstrates the hierarchy of federated ledger fabric, in which a inter-ledger consensus network acts as a hub to interconnect multiple intra-ledger consensus networks. For a intra-ledger network, domain administrator selects a subset of the nodes as a validator committee, which executes an efficient BFT consensus protocol to maintain a private ledger of the domain. Given assumption that no more than $f$ validators within a committee are byzantine ones, BFT consensus ensures the ultimate goal of agreement if a committee includes $n \geq 3f+1$ total validators. The validators of committee collect transactions sent by nodes of intra-ledger network and commit them on a private ledger managed by the domain. Enforcing a BFT consensus on a small-scale validator committee can reduce messages propagation delay and communication cost in the domain network, such that high throughput and low latency of committing transactions are guaranteed.

The whole system includes multiple fragmented domain networks, and each domain has varying account address management and private ledger status. Thus, nodes in different intra-ledger network do not directly communicate with each other. We use a PoW-based inter-ledger consensus network as a public trust-free intermediate to enable security and privacy preservation of inter-ledger transactions across heterogeneous private intra-ledger networks. Each domain administrator specifies a set of resource-rich nodes as miners to join the public inter-ledger network. The scalable PoW consensus ensures a probabilistic finality on inter-ledger transactions if majority (51\%) of miners are honest. 

For cross-domain operations, nodes within an intra-ledger network rely on a set of miners to commit inter-ledger transactions on the inter-ledger network. Unlike the intra-ledger transaction that is only propagated and valid in the domain network, a inter-ledger transaction also includes the address of a inter-ledger miner in its network. Then, the miner records an inter-ledger checkpoint on the inter-ledger as a public proof, which includes the proof of intra-ledger transaction along with private ledger summary. All data on public inter-ledger are transparent such that all miners can query checkpoints from the inter-ledger. 

The inter-domain transactions can be mutual verified in cross-ledger operations, as Fig. \ref{fig:Fed-DDM_arch} shows. A node can verify checkpoints of intra-ledger transactions from the different domain networks through the help of a miner within its domain. Only checkpoints are stored on the public inter-ledger, but raw transactions and intra-ledger status are protected by access control and privacy policies of the domain network. Therefore, sensitive information are prevented from being exposed on public inter-ledger network.

\section{Inter-ledger Transaction Protocol}
\label{sec:crossledger}
Inter-ledger transactions involve sellers and buyers from different domains without fully trust. Therefore, trust-free, auditability and traceability become critical in cross-domain data services. Each domain has full controls of data on its private ledger with different access control and privacy policies. Thus, trade-offs must be addressed between privacy-preservation of intra-ledger and transparency of inter-ledger. Fed-DDM adopts a hybrid intra-ledger \& inter-ledger data structure and designs a smart contract enabled inter-ledger protocol to support secure and privacy-preserving inter-ledger transactions.    

\subsection{Inter-ledger Transaction Data Structure}

\begin{figure} [t]
\begin{center}
\begin{tabular}{c}
\includegraphics[height=5.8 cm]{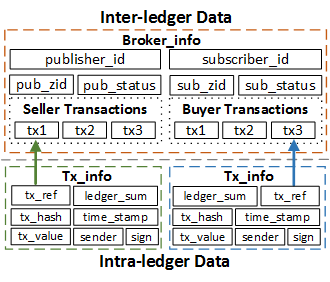}
\end{tabular}
\end{center}
\vspace{-15pt}
\caption[example] { \label{fig:inter-ledger_data} Inter-ledger transaction data structure.}
\vspace{-15pt}
\end{figure}

To accomplish cross-domain operations, a seller or a buyer launches a set of sequential intra-ledger transactions within its private network, then he/she can choose a publisher or subscriber of the same domain to commit inter-ledger transactions on the public federated ledger. As publishers and subscribers also have valid blockchain accounts on the inter-ledger network, they can delegate their sellers or buyer to record checkpoints of intra-ledger transactions on the inter-ledger for further cross-domain data services. 

Figure \ref{fig:inter-ledger_data} illustrates a hybrid intra-ledger \& inter-ledger data structure for cross-domain data service transactions. \emph{Broker\_info} is the basic data unit of inter-ledger transaction storage, which includes participants' information and data service processing status. $publisher\_id$ and $subscriber\_id$ store account addresses of publisher and subscriber of inter-ledger network. $pub\_zid$ and $sub\_zid$ save their intra-ledger network IDs, which are also called as zone IDs. For each transaction sent by a seller or buyer, the $tx$ only stores a reference $tx\_ref$ pointed to the raw transaction $Tx\_info$, which is recorded on the private intra-ledger. 

As $tx\_ref$ is a hash value with fixed length of 32 or 64 bytes, each $tx$ has the same size even if the linked raw data $tx\_value$ saves a large value or requires a complicated format. It reduces the ever-increasing data storage overhead by recording references as checkpoints rather than raw transactions on the inter-ledger. Moreover, the $Tx\_info$ can only be accessed by authorized participants in data service exchanges. Thus, other nodes outside the domain cannot gain transaction details given solo $tx$ on public inter-ledger, and privacy-preservation of intra-ledger is guaranteed.  

\subsection{Smart Contract enabled Inter-ledger Protocol}

\begin{figure} [t]
\begin{center}
\begin{tabular}{c}
\includegraphics[height=6.25 cm]{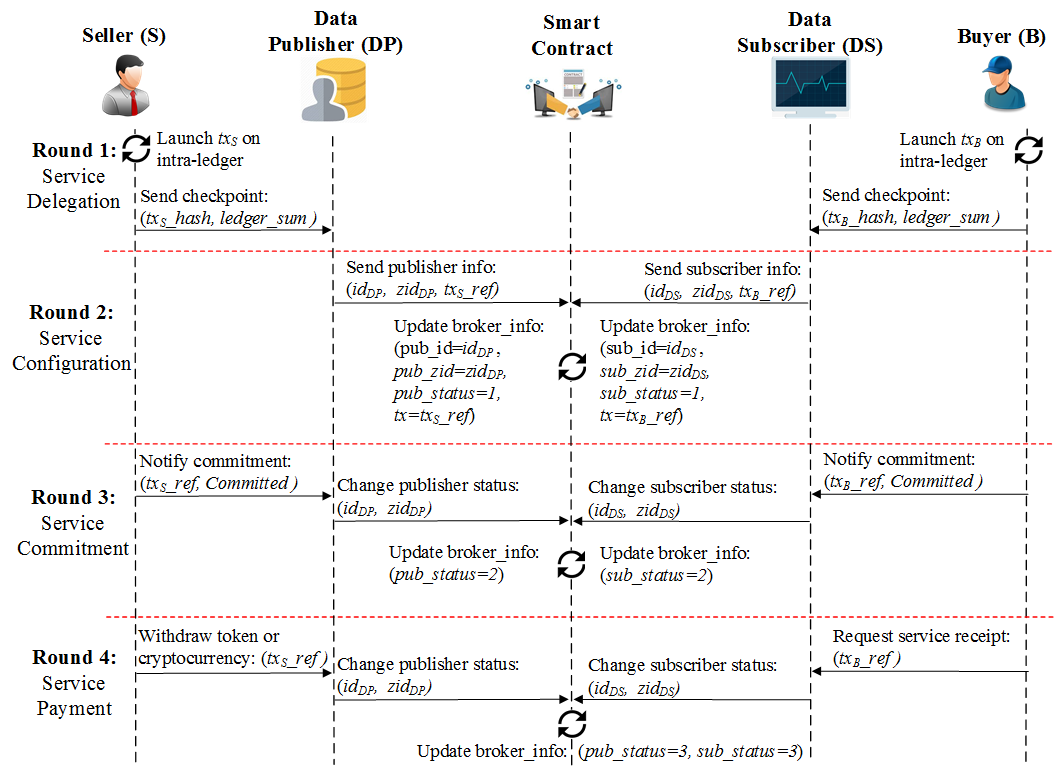}
\end{tabular}
\end{center}
\vspace{-10pt}
\caption[example] { \label{fig:inter-ledger_protocol} Inter-ledger transaction protocol.}
\vspace{-15pt}
\end{figure}

Leveraging a federated access control strategy \cite{xu2018federated}, the system administrator of each domain can deploy data service exchange smart contracts (SCs) on the inter-ledger network, and only authorized miners are allowed to update SCs. The SC utilizes a publish-subscribe service agreement mode to accomplish inter-ledger transactions among participants from different domains. Figure \ref{fig:inter-ledger_protocol} demonstrates a high-level view of inter-ledger transaction protocol, and key procedures of workflows are described as follows.

\subsubsection{Service Delegation}
At the beginning of a data service exchange task, seller (buyer) must launch a intra-transaction $tx$ to record the requirements, like data descriptions, access privileges and payment conditions, etc.. Until the $tx$ has been committed on the intra-ledger, seller (buyer) sends a delegation request including checkpoint data of $tx$ to any registered data publisher (DP) (data subscriber (DS)) within the domain. After successfully verifying identity of seller (buyer) as well as its $tx$ and ledger summary, DP (DS) sends back acknowledgement. Otherwise, a deny notification is returned.

Given an assumption that DP and DS are honest, only the crash fault scenario that original DP (DS) is off-line is considered. As a sequential delegation list is used by the federated access control, a seller (buyer) can choose next available miners in the delegation list as DP (DS) if the original DP (DS) fails to serve operations. Then the system administrator can update broker information of SC with new DP (DS) accordingly. This rule is enforced during the entire lifetime of the inter-ledger transaction protocol.

\subsubsection{Service Configuration}
As DP and DS are also authorized miners on the inter-ledger network, they can update the the broker information on SCs to finish service configuration. The service configuration allows DP (DS) to delegate a seller (buyer) for data service proposal and negotiation. If no SC can meet the requirements, DP (DS) chooses an empty SC to propose a sell (purchase) request. Otherwise, DP (DS) selects an optimal SC with initial broker information, then updates publisher (subscriber) related data fields of SC accordingly. Because DS is the actual payee of $tx$ sent by buyer, DS also deposits token or cryptocurrency from the buyer into the account of SC, which acts as a trust-free intermediate to support a cross-domain payment. Finally, the broker data of SC are sent back to the seller (buyer) as checkpoints, which are used for cross-domain operations.  

\subsubsection{Service Commitment}
Service commitment round is triggered only if the seller has delivered data service and the buyer is ready for payment. The seller (buyer) simply sends a notification to DP (DS) for service commitment. If both $pub\_status$ and $sub\_status$ of SC are $1$, then DP (DS) changes the broker's status to \emph{committed} and sends back a response. Otherwise, DP (DS) denies the service commitment by returning error messages to the seller (buyer). Only if both DP and DS have successfully changed the broker status to committed, procedures of data service protocol can move forward to the payment round.  

\subsubsection{Service Payment}
A service payment starts if a seller sends a withdraw request to DP or a buyer retrieves a service receipt from DS. After receiving requests, DP (or DS) changes broker status on SC, which also automatically triggers procedure of transferring balance of SC account to DP. Thanks to atomicity of procedures in SC, secure inter-ledger balance transfer and correct broker status update are guaranteed. Because DP is the actual payer of $tx$ sent by seller, token or cryptocurrency will be transferred from DP to seller within the domain network.      

\section{Experiment and Evaluation}
\label{sec:experiment}
\subsection{Prototype Implementation}
To verify the proposed Fed-DDM scheme, a proof-of-concept prototype is implemented in Python. We adopt Flask \cite{flask}, which is a light micro-framework for web-service application, to expose a set of RESTful APIs for FEd-DDM applications. All cryptographic functions are developed on the foundation of standard python lib cryptography \cite{pyca}, like using RSA for key generation and digital signature, and using SHA-256 for all hash operations. To simulate inter-ledger transactions based on a hierarchical blockchain system, we use Ethereum \cite{ehtereum} to setup a inter-ledger network, and two intra-ledger networks are built on Tendermint core \cite{tendercore}. We use Solidity \cite{solidity}, which is a contract-oriented and high-level language, to develop smart contracts.

Table \ref{tab:testbed} shows configuration of devices. For a private Ethereum network, six miners are deployed on six separate desktops and other devices work as non-miners. All devices use Go-Ethereum \cite{goethereum} as client applications to interact with Ethereum network. 
Each Tendermint enabled intra-ledger network includes 10 validators, and each validator is hosted on a Raspberry Pi (RPi). All devices in testbed are connected through a local area network (LAN). 

\begin{table}[ht]
\vspace{-0.10in}
\caption{Configuration of Experimental Devices.} 
\vspace{-0.18in}
\label{tab:testbed}
\begin{center}       
\begin{tabular}{|l|p{3.0cm}|p{3.0cm}|} 
\hline
\rule[-1ex]{0pt}{3.5ex} \textbf{Device} & Dell Optiplex-7010 & Raspberry Pi 4 Model B \\
\hline
\rule[-1ex]{0pt}{3.5ex} \textbf{CPU} & Intel Core TM i5-3470 (4 cores), 3.2GHz & Broadcom ARM Cortex A72 (ARMv8), 1.5GHz \\
\hline
\rule[-1ex]{0pt}{3.5ex} \textbf{Memory} & 8GB DDR3 & 4GB SDRAM \\
\hline
\rule[-1ex]{0pt}{3.5ex} \textbf{Storage} & 350G HHD & 64GB (microSD card) \\
\hline
\rule[-1ex]{0pt}{3.5ex} \textbf{OS} & Ubuntu 16.04 & Raspbian GNU/Linux (Jessie) \\
\hline
\end{tabular}
\end{center}
\vspace{-0.20in}
\end{table}

\subsection{Performance Evaluation}
To evaluate the performance of the FEd-DDM prototype, key security functions, like identity authentication, access control, broker and transaction verification, are encapsulated into multiple containerized microservices units. The four types of security services are deployed on four desktops and four RPis separately, and each machine only runs a single containerized service unit. The cost of message encryption and decryption are not considered during the test.

\subsubsection{Computation Overhead}

\begin{figure} [t]
\begin{center}
\begin{tabular}{c}
\includegraphics[height=4.3 cm]{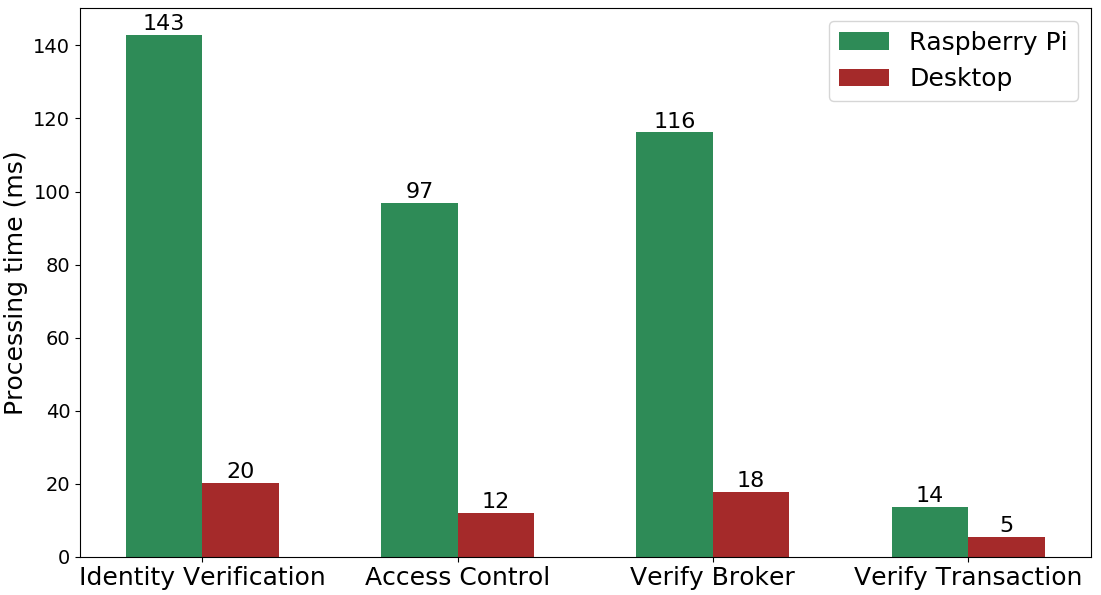}
\end{tabular}
\end{center}
\vspace{-10pt}
\caption[example] { \label{fig:process_latency} Processing time of security services on different platforms.}
\vspace{-15pt}
\end{figure}

To evaluate computation overhead of security services on host machines, one-hundred Monte Carlo test runs have been done in total given test scenario that a client sends one transaction per second (TPS) to a microservice unit. Figure \ref{fig:process_latency} shows average processing time of security services in Fed-DDM on desktop and RPi benchmarks. The verifying raw transaction is performed by authorized nodes of a domain network, and it relies on a set of efficient Application BlockChain Interface (ABCI) \cite{tendercore} to query intra-ledger transactions from a private Tendermint network. Given execution on the same machine, verifying transaction requires less processing time than SC enabled security services. Thus, it is affordable on resource limited IoT devices working on the intra-ledger network.  

Owing to high computation resource required by smart contract operations on Ethereum Virtual Machine (EVM) \cite{ehtereum}, running SC enabled security services on RPi incurs much more time latency than executing the same tasks on desktop. However, miners and non-miners working on inter-ledger network are powerful devices, like desktops or edge servers, which only bring limited overhead (about 20 ms). Compared with calculated average network delay of delivering messages (about 200 ms), time latency incurred by security services is acceptable in the inter-ledger environments. 

\subsubsection{Transaction Latency}

\begin{figure} [t]
\begin{center}
\begin{tabular}{c}
\includegraphics[height=4.0 cm]{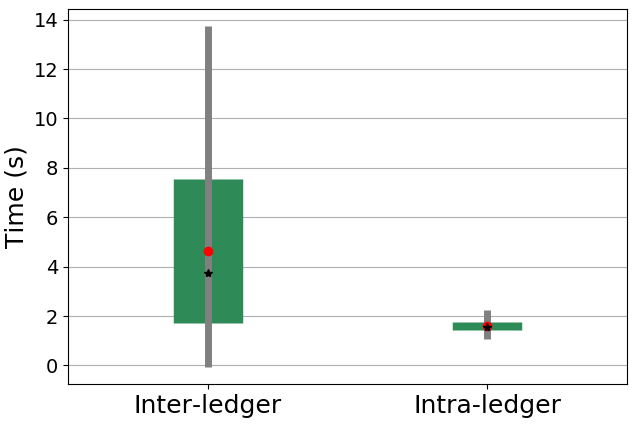}
\end{tabular}
\end{center}
\vspace{-10pt}
\caption[example] { \label{fig:tx_latency} Time latency for committing transactions.}
\vspace{-15pt}
\end{figure}

As a key performance metric of blockchain network, transaction ($tx$) committed time is closely related to block confirmation time influenced by the consensus protocol. For the inter-ledger network, a desktop simulates DP (DS) launches transactions to update broker information on Ethereum. Regarding intra-ledger network, a RPi simulates seller (buyer) which records intra-ledger transactions on Tendermint. We conducted 100 Monte Carlo test runs and evaluated the end-to-end time latency that is how long a $tx$ can be finalized on its blockchain. 

Figure \ref{fig:tx_latency} shows the distributions of time delay for committing transactions given different blockchain networks. Each green bar indicates standard deviation (std) with a mean represented by red dot, and the gray line shows whole data range with the black star as median point. Thanks to an efficient BFT consensus protocol used by Tendermint, mean of intra-ledger $tx$ committed time is 1.6 s with small deviation (0.146 std). Such a low and stable transaction latency is suitable for time sensitive scenarios in domain network. Ethereum relies on a probabilistic PoW consensus, which has variable block confirmation time. Thus, $tx$ committed time in inter-ledger network is varying with large deviation (2.926 std). As inter-ledger is mainly to guarantee auditability and payment for corss-domain operations, and it does not need strong synchronization and consistency required by intra-ledger. Therefore, 4.5 s mean of $tx$ committed time is acceptable for inter-ledger transactions.

\subsubsection{Comparative Evaluation}

Table \ref{tab:comparison} provides a comparison of running Fed-DDM system by committing inter-ledger and intra-ledger transactions. We calculate $tx$ rate $tx/s$ to evaluate transactions throughput, which indicates how many $tx$ can be processed per second. Ethereum block size is bounded by how many units of gas can be spent per block, which is known as the block gas limit~\cite{ethblocksize}. Given the maximum block size (12,000,000 Gas) and the base cost of any transaction (21,000 Gas, that is accessed at July 20, 2020), each block in Ethereum can include around 571 transactions. Thus, $tx$ rate is (571/4.5)$\approx$126 $tx/s$ under inter-ledger network.
Tendermint specifies the fixed 1MB block size and 1 KB per transaction such that a block can store the maximum of 1000 transactions. Therefore, $tx$ rate of intra-ledger is (1000/1.6)=625 $tx/s$, which is promising to meet high throughput requirement of domain networks.

Regarding resource consumption, we use ``top'' command to monitor CPU and memory usage on desktop (Ethereum miner) and RPi (Tendermint validator). The computation intensive PoW mining algorithm used by Ethereum almost occupies full CPU capacity and consumes about 1.2GB memory. Therefore, huge computation cost prevents resource constrained IoT devices from running as miners in inter-ledger network. Unlike Ethereum, Tendermint uses a lightweight BFT consensus algorithm to achieve efficiency in CPU and memory usage. Thus, it is suitable for deploying validators on IoT devices to maintain intra-ledger network.

In Ethereum networks, a transaction commitment requires gas that is paid to miners as rewarding. The average gas fee for each transaction is 0.001 Ether, which amounts to \$1.23 given the Ether price in the public Ethereum market (\$1233.67/Ether at Jan 22, 2021). Compared with Ethereum, Tendermint is a permissioned network without requiring transaction fees that are associated with any type of currency. Therefore, intra-ledger transactions do not introduce additional financial cost. Each domain relies on its own incentive mechanism to encourage more participants to join intra-ledger network and gain benefits.

\begin{table}[t]
\caption{Comparative evaluation of running Fed-DDM.} 
\vspace{-0.18in}
\label{tab:comparison}
\begin{center}       
\begin{tabular}{|l|c|c|} 
\hline
\rule[-1ex]{0pt}{3.5ex}  &  Inter-ledger & Intra-ledger \\
\hline
\rule[-1ex]{0pt}{3.5ex} \textbf{$tx$ rate (tx/s)} &  126  &  625 \\
\hline
\rule[-1ex]{0pt}{3.5ex} \textbf{CPU usage (\%)} &  100  &  32 \\
\hline
\rule[-1ex]{0pt}{3.5ex} \textbf{Memory usage (MB)} &  1,200  &  70 \\
\hline
\rule[-1ex]{0pt}{3.5ex} \textbf{Gas/$tx$ (Ether) } &  0.001  &  $\times$ \\
\hline
\end{tabular}
\end{center}
\vspace{-0.20in}
\end{table}

\subsection{Security Analysis}
We assume that an adversary has limited capacity and is subject to the usual cryptographic hardness guarantees. Thus, the security properties of the Fed-DDM system relies on the security of hybrid intra-ledger and inter-ledger consensus network. Given assumption that an adversary could control no more than $f$ byzantine validators in an intra-ledger network including total $n \geq 3f+1$ validators, BFT consensus protocol can guarantee \emph{safety}, which requires all $2f+1$ honest validators to agree on a same total order of blocks on local distributed ledger. As BFT leverages voting procedures to finalize blocks in synchronous network environments, a \emph{liveness} property is achieved by committing all valid transactions within the consensus round. As a trade-off, relying on a pre-fixed validators committee to maintain intra-ledger is inevitable to reduce security. Using a committee randomness security scheme \cite{microchain21} is promising to ensure unpredictability of committee election, and we leave it for future work. 

In a public inter-ledger network, if majority (51\%) of miners are honest and they can correctly execute Nakamoto consensus protocol, the safety is guaranteed through a probabilistic finality on blocks on inter-ledger. Once checkpoints and transactions are finalized on immutable public inter-ledger, it is hard for an adversary to launch double spending attacks by reverting transactions or status of smart contracts. In addition, an inter-ledger transaction only includes publisher (subscriber) information along with checkpoints of raw intra-ledger transactions within different domain networks. Thus, attackers cannot track sellers (buyers) during data service by solo interlinking transparent transactions recorded on public inter-ledger. Furthermore, each domain relies on fine grained access control and privacy policies to protect data on its intra-ledger. Only authorized participants of a inter-ledger transaction, like seller and user, are allowed to access linked raw intra-ledger transactions and ledger summary for cross-domain operations. Therefore, other sensitive information on intra-ledger is protected.

\section{Conclusions}
\label{sec:conclusion}
This paper introduces Fed-DDM, a secure-by-design and federated ledgers based blockchain framework, which aims to enhance security and privacy-preserving properties for hierarchical decentralized data marketplaces. Combining federated security networking framework and hybrid intra-ledger and inter-ledger consensus protocols, Fed-DDM improves scalability and performance trade-offs in mono-chain network, and ensures security and privacy preservation for
inter-domain operations. The experimental results based on a proof-of-concept prototype are very encouraging. 

However, some open questions are yet to be solved before applying Fed-DDM in the real-world, like security of small scale intra-ledger consensus network and incentive mechanism for inter-ledger transactions across different domains. Our on-going efforts include validating the proposed architecture in a real-life data-driven DDM system, and ensuring overall system-design stability, efficiency and effectiveness. This includes a fully functional proof-of-concept prototype implementation and a comprehensive analysis of assessment of security features.

\bibliographystyle{IEEEtranS} 
\bibliography{report} 
\end{document}